\documentclass{article}
\usepackage[utf8]{inputenc}
\usepackage{main}
\usepackage{microtype}
\usepackage{graphicx}
\usepackage{times}
\usepackage{amsmath}
\usepackage{amssymb}
\usepackage{enumitem}
\usepackage{booktabs}
\usepackage{xcolor}     
\usepackage{natbib}
\definecolor{mydarkblue}{rgb}{0,0.08,0.45}
\usepackage[colorlinks=true,linkcolor=mydarkblue,citecolor=mydarkblue,filecolor=mydarkblue,urlcolor=mydarkblue]{hyperref}
\usepackage{fancyhdr}

\definecolor{gred}{RGB}{250, 210, 207}
\definecolor{coolblue1}{rgb}{0.91, 0.94, 0.98}
\definecolor{coolblue2}{rgb}{0.76, 0.85, 0.94}
\definecolor{coolblue3}{rgb}{0.54, 0.72, 0.87}
\definecolor{coolblue4}{rgb}{1, 1, 1}

\usepackage{xspace}
\usepackage{cleveref}

\usepackage[textsize=tiny]{todonotes}

\usepackage{multirow}
\usepackage{xspace}
\usepackage{wrapfig}
\usepackage{enumitem}
\usepackage{makecell}
\usepackage{caption}
\usepackage{subcaption}
\usepackage[table]{xcolor}

\usepackage[most]{tcolorbox}
\usepackage{etoolbox}
\tcbuselibrary{listingsutf8}

\usepackage{tabularx}
\newcolumntype{L}{>{\raggedright\arraybackslash}X}

\newtcolorbox{LLMPrompt}[1][]{
  enhanced, breakable, verbatim,
  colback=gray!5, colframe=black!15, boxrule=0.4pt, arc=1.5mm,
  left=6pt,right=6pt,top=6pt,bottom=6pt,
  fontupper=\small\ttfamily,
  colbacktitle=gray!50,
  title=#1
}

\newcommand{\correct}{FuncPass\xspace}
\newcommand{\secu}{SecPass\xspace}
\newcommand{\method}{\textsc{SecureVibe}\xspace}

\definecolor{mediumgreen}{RGB}{0, 150, 50}
\definecolor{mediumred}{RGB}{180, 30, 30}

\newcommand{\improve}[1]{\textcolor{mediumgreen}{#1}}
\newcommand{\dropperf}[1]{\textcolor{mediumred}{#1}}

\usepackage{lineno}

\definecolor{darkblue}{rgb}{0, 0, 0.5}
\hypersetup{colorlinks=true, citecolor=darkblue, linkcolor=darkblue, urlcolor=darkblue}

\title{SecureVibe: Making Vibe Coding More Secure}

\author{
\textbf{Danqing Wang}$^{1,2}$\thanks{This work was done while
Danqing Wang, Zhepei Wei, and Isadora White were interns
at Microsoft Research.} \quad
\textbf{Baolin Peng}$^{2}$ \quad
\textbf{Zhepei Wei}$^{2,3}$ \quad
\textbf{Isadora White}$^{2,4}$ \quad
\textbf{Wenlin Yao}$^{2}$ \quad
\textbf{Hao Cheng}$^{2}$ \\[0.2em]
\textbf{Qianhui Wu}$^{2}$ \quad
\textbf{Minseon Kim}$^{2}$ \quad
\textbf{Xingdi Yuan}$^{2}$ \quad
\textbf{Lei Li}$^{1}$ \quad
\textbf{Jianfeng Gao}$^{2}$ \\[0.6em]
\textsuperscript{1}Carnegie Mellon University \quad
\textsuperscript{2}Microsoft Research \\
\textsuperscript{3}University of Virginia, Charlottesville \quad
\textsuperscript{4}University of California, San Diego \\
\href{https://github.com/MSR-Orchard/SecureVibe}{\includegraphics[height=0.4cm]{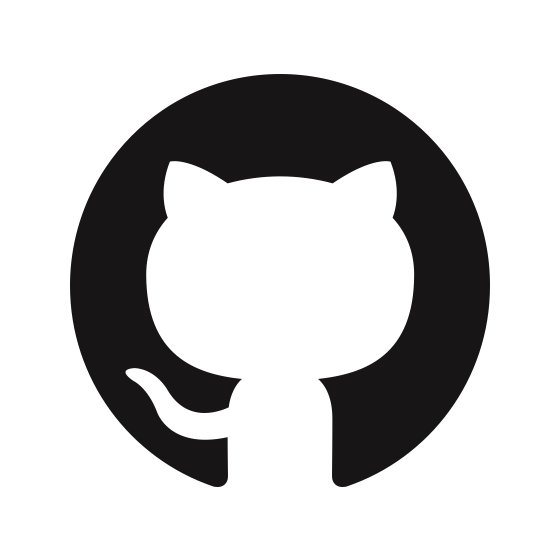} \textbf{MSR-Orchard/SecureVibe}} ~ ~ ~ \href{https://huggingface.co/datasets/dqwang122/SafeVibe}{\includegraphics[height=0.4cm]{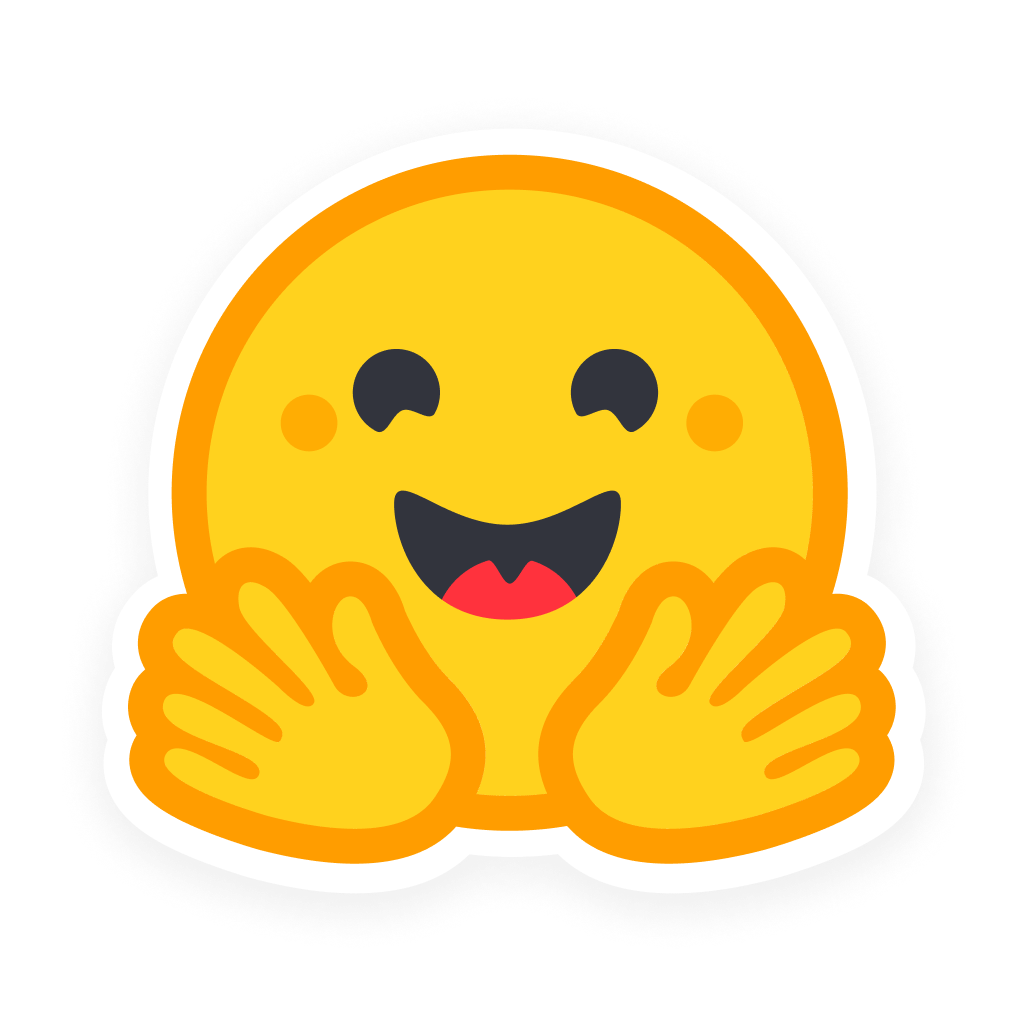} \textbf{dqwang122/SafeVibe}}
}

\begin{document}

\maketitle
\thispagestyle{fancy}
\fancyhead{}
\lhead{\includegraphics[height=0.5cm]{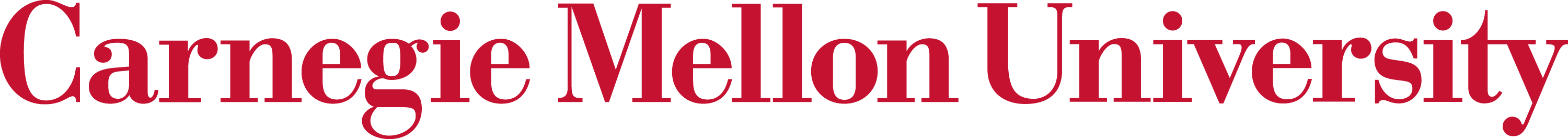}}
\chead{\includegraphics[height=0.5cm,trim=0 670bp 0 670bp,clip]{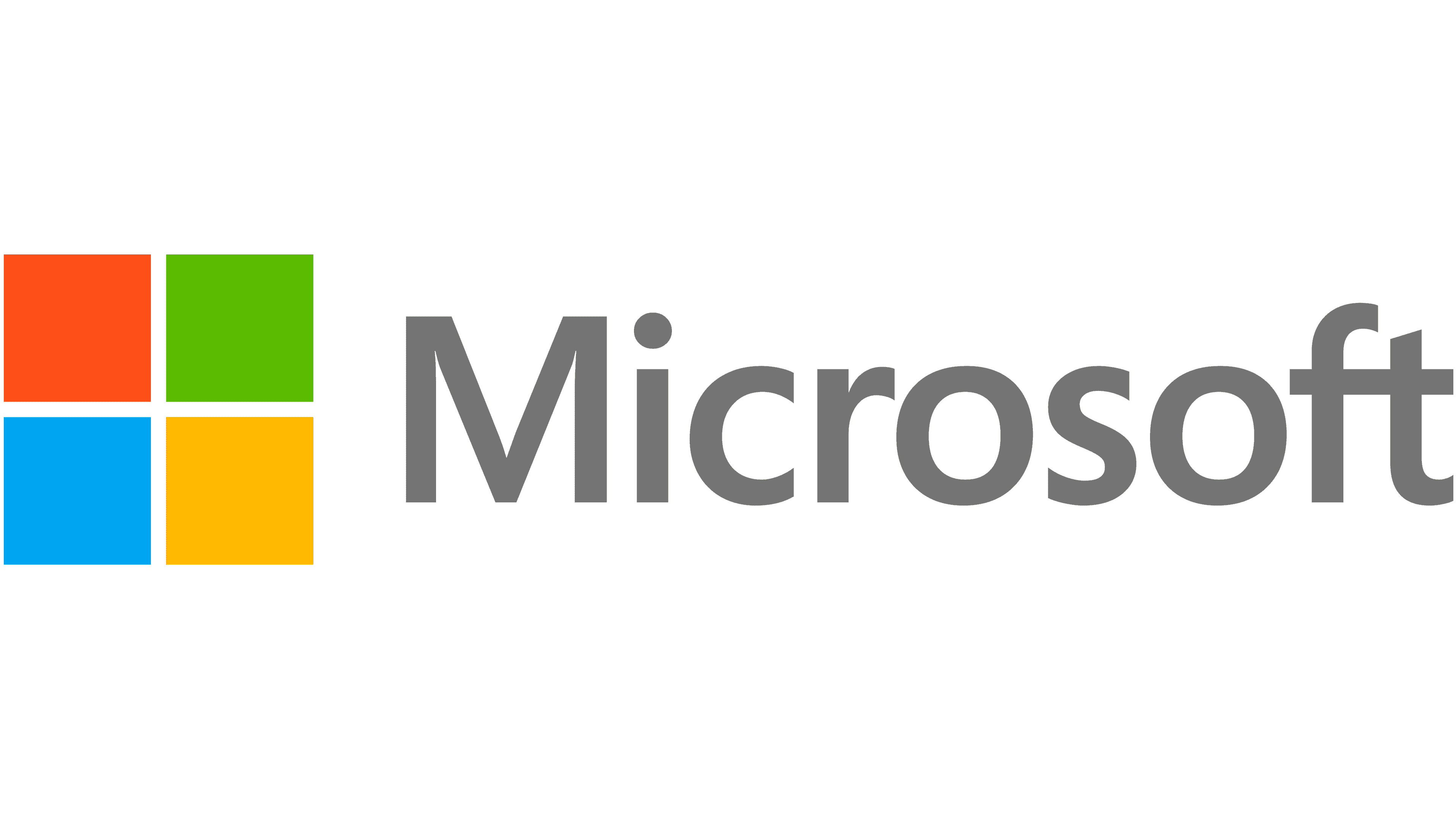}}
\rhead{%
  \raisebox{-0.1cm}{\includegraphics[height=0.8cm]{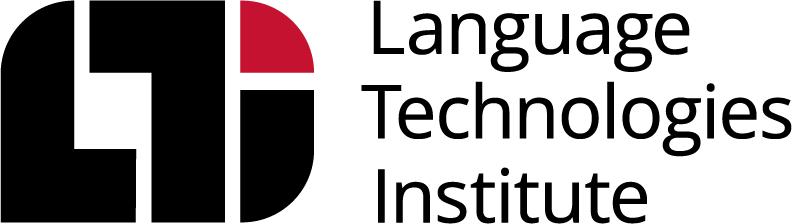}}%
}
\renewcommand{\headrulewidth}{0pt}
\setlength{\headheight}{12pt}
\addtolength{\topmargin}{00pt}
\setlength{\headsep}{3mm}

\vspace{-1.0em}

\begin{abstract}
As vibe coding becomes increasingly capable and widespread, security vulnerabilities in even functionally correct solutions are a growing concern. When investigating functionally correct but insecure solutions, we find that the insecure agent is less than half as likely to conduct effective planning and testing for the hidden security risks behind the functional requirements. 
Motivated by this, we develop \method, a training recipe that explicitly targets planning and testing for code security. \method constructs training signals around these security behaviors. It includes supervised fine-tuning on the security suite with 4 security tasks, and post-training methods $\method_{rl}$ and $\method_{hg}$ to enhance security capabilities from verifiable execution feedback and hint-based self-supervision. 
Our \method outperforms the baseline on two types of security coding tasks across 4 benchmarks. Specifically, \method improves the security pass@1 by 6.9 points on BaxBench. The gains extend to unseen CWE categories, with improvements of 11.5 points on SusVibes. Meanwhile, it also improves functionality pass@1 by 13.6 points on the security coding task SusVibes and 4.1 points on the generic coding task SWE-bench Verified. 
Further analysis offers two practical insights: (i) diversifying supervision across security planning, coding, and testing strengthens security behaviors more effectively than adding coding trajectories alone, and (ii) hint-guided supervision is particularly valuable when the agent’s existing security capabilities are insufficient to learn effectively from outcome feedback.

\end{abstract}

\section{Introduction}
\label{sec:intro}
\begin{figure}[ht]
    \centering
    \includegraphics[width=\linewidth]{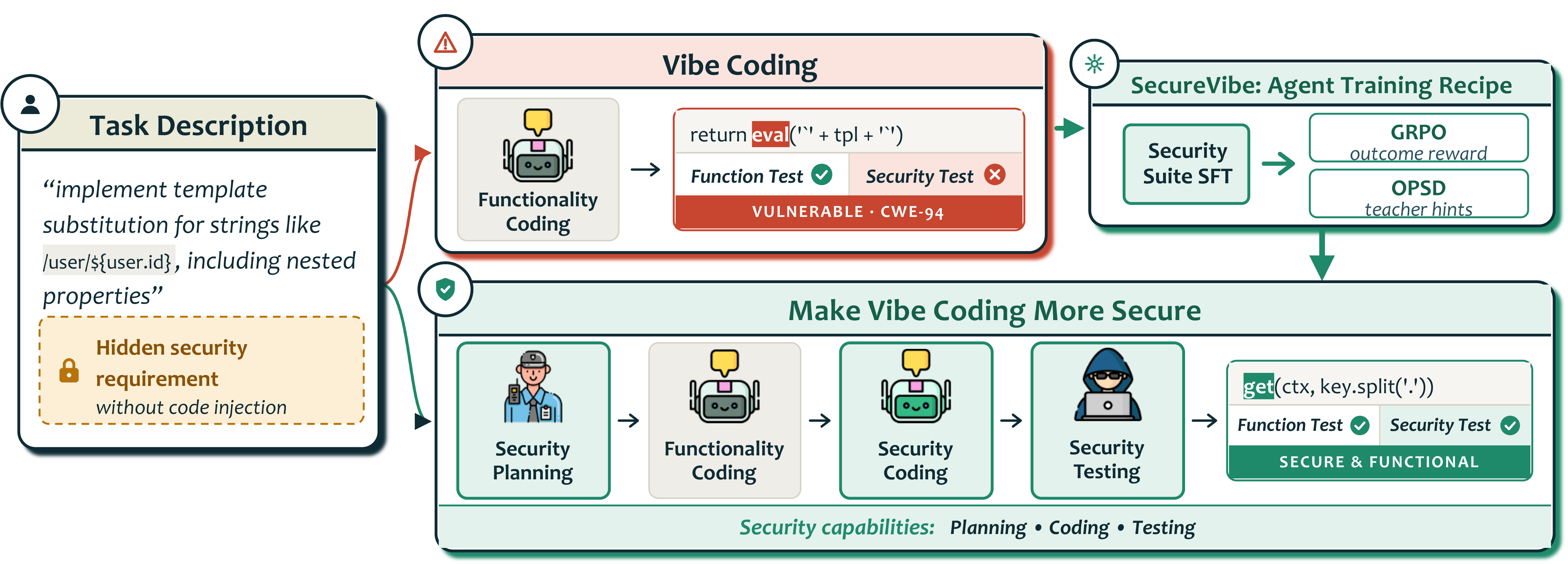}
    \caption{\method: making vibe coding more secure by enhancing security capabilities via a mixture of security tasks and post-training. }
    \label{fig:safevibe}
\end{figure}

Large language model (LLM) coding agents can now modify real-world software repositories with increasing success. Given a natural-language request, these agents can inspect a codebase, plan changes, edit multiple files, and execute tests. These capabilities have made ``vibe coding'' an increasingly practical workflow in which developers describe the desired behavior and rely on an agent to implement it. Progress on benchmarks such as SWE-bench~\citep{jimenezswe} and Terminal-Bench~\citep{merrill2026terminal} further shows that coding agents can solve complex software-engineering tasks beyond isolated code completion.

However, functional correctness does not always guarantee security. An implementation may pass its functional tests while still introducing vulnerabilities~\citep{ye2026coding,baumann2026swe}. These vulnerabilities may result from unsafe input handling, missing authorization checks, insecure API usage, or other security-sensitive decisions. This risk is especially important in vibe coding because users may accept an agent's changes without carefully auditing them. Existing approaches mainly improve functional task completion through better agent scaffolding, supervised fine-tuning, or reinforcement learning~\citep{pantraining,ma2024lingma,wei2025swerl}. Secure-code benchmarks usually measure whether generated code contains vulnerabilities~\citep{vero2025baxbench,zhao2026is,chen2025secureagentbenchbenchmarkingsecurecode}. However, they rarely explain what capabilities coding agents lack or how agents can acquire them. Therefore, it remains unclear why functionally capable agents produce insecure code and how training can jointly improve secure and general coding capabilities.

We study secure vibe coding through three research questions. \textit{RQ1: What capabilities make a coding agent secure?} \textit{RQ2: How can an agent acquire these capabilities?} \textit{RQ3: Do these capabilities generalize to new security settings?} 

For RQ1, we compare the trajectories between the insecure and secure implementations. We find that the insecure agent, although it also provides a functional solution, is less than half as likely to plan and test for the hidden security requirements. Such a lack of security awareness makes them stop after passing functionality tests, ignoring the security risks.

Motivated by these findings, we develop \method, a security training recipe covering data-mixture construction and model training to answer RQ2. As shown in \autoref{fig:safevibe}, $\method_{base}$ combines \textit{Functionality-Focused Coding, Security-Focused Coding, Security Planning, and Security Testing} to teach agents to recognize security requirements, implement protective controls, and synthesize security tests. We then design \textbf{outcome-based $\method_{rl}$} and \textbf{hint-guided $\method_{hg}$} to investigate the effectiveness of further post-training. 
$\method_{rl}$ learns from executable functional and security feedback, while $\method_{hg}$ receives dense token-level supervision from a self-distillation teacher equipped with instance-specific security hints.

To answer RQ3, we train Qwen 3.5 35B with \method on AutoBaxBench~\citep{arx2026autobaxbuilder} and PatchEval-Gen~\citep{wei2025patcheval} and evaluate its effectiveness on two more benchmarks SusVibes~\citep{zhao2026is} and BaxBench~\citep{vero2025baxbench}. PatchEval-Gen and SusVibes require coding agents to implement features in real-world repositories, while AutoBaxBench and BaxBench ask the agent to implement the repo-level backend from scratch. 
In the validation set of PatchEval-Gen and AutoBaxBench, $\method_{rl}$ improves average \correct by 9.78 percentage points, while $\method_{hg}$ improves average \secu by 8.23 percentage points over the base model. These gains generalize to SusVibes and BaxBench, where the corresponding improvements reach 10.9 and 4.8 points. On unseen-CWE subsets, \method improves \secu by 11.5 points on SusVibes and 7.1 points on BaxBench. Security training also strengthens general coding performance: every trained checkpoint outperforms the base model on SWE-bench Verified~\citep{jimenezswe}, with gains of up to 4.1 points.

Further analysis shows that \method elicits planning, coding, and testing for security more frequently than the other recipes, behaviors we identified as essential for secure solutions. It also provides practical guidance for post-training: first, trigger the agent's security behaviors by supervised training with mixed security-related tasks. Then, choose the appropriate post-training methods to further enhance security: use $\method_{rl}$ if the SFT checkpoint always shows great security capabilities for dense positive outcome feedback; otherwise, use $\method_{hg}$ with hint-based teacher guidance to get denser supervision.

Our main contributions are as follows:
\begin{itemize}[leftmargin=*,nosep]
    \item We identify that insecure agents usually lack awareness to plan and test for the implicit security requirements, leading to a performance gap between functionality and security (Section~\ref{sec:motivation}).
    \item We introduce \method, a secure-coding training recipe that combines Security Suite SFT with outcome-based $\method_{rl}$ or hint-guided $\method_{hg}$ (Sections~\ref{sec:sft} and~\ref{sec:post}).
    \item We demonstrate functional and security gains under held-out and cross-benchmark evaluation, generalize to unseen CWE categories, and improve general coding performance (Section~\ref{sec:experiments}).
    \item We analyze the complementary strengths of $\method_{rl}$ and $\method_{hg}$ and derive practical guidance for selecting post-training methods based on reward density (Section~\ref{sec:analysis}).
\end{itemize}

\section{Related Work}
\label{sec:related}

\textbf{Coding Agents and Optimization.}
LLM-based coding agents can perform repository-level tasks such as bug fixing, feature implementation, test generation, environment setup, and library generation~\citep{jimenezswe,mundler2024swt,eliseeva2025envbench,zhaocommit0}. Prior work improves these agents through better actions, workflows, and inference-time scaling~\citep{yang2024swe,xia2025demystifying,gao2025trae}, or through supervised and reinforcement-based model training~\citep{zhang2025sealign,ma2025sorft,wei2025swerl,kim2026frognano}. These methods primarily optimize functional task completion rather than implementation security.
Among general post-training methods, GRPO estimates advantages from relative rewards within groups of responses without training a separate critic~\citep{shao2024deepseekmath}. DAPO improves GRPO through dynamic sampling and asymmetric clipping~\citep{yu2026dapo}, while GSPO moves importance weighting and clipping from the token level to the sequence level~\citep{zheng2025gspo}. By contrast, OPD learns from teacher feedback on trajectories sampled by the student~\citep{song2026survey}, and OPSD uses the same model as both teacher and student under different contexts~\citep{zhao2026selfdistilled}. These two families motivate our comparison of reward-based optimization and self-distillation for security-oriented coding agents.

\textbf{Secure-Code Benchmarks.}
Various benchmarks have emerged to assess both the security and functional correctness of LLM-generated code.
Earlier benchmarks focused on single-turn generation in limited settings, such as a single file or function~\citep{siddiq2024sallm,peng2025cweval,yang2024seccodeplt,pearce2025asleep}.
More recent benchmarks have expanded their scope to repository-level tasks that may require edits across multiple files.
BaxBench~\citep{vero2025baxbench} and AutoBaxBuilder~\citep{arx2026autobaxbuilder} evaluate backend application security across popular frameworks and programming languages using functional tests, security tests, and expert-designed exploits.
SecureAgentBench~\citep{chen2025secureagentbenchbenchmarkingsecurecode} mine repository-level vulnerability-fixing commits and repurpose them as benchmark tasks. SusVibes~\citep{zhao2026is} investigates the gap between functional correctness and security across vibe coding agent frameworks. These benchmarks focus primarily on evaluation and provide limited data for training security-aware coding agents.

\textbf{Secure Code Generation.}
Prior work improves secure code generation through supervised tuning on secure and insecure examples and reinforcement learning with synthesized security tasks and multi-objective rewards~\citep{he2024instruction,liu2026purpcode}. Related functional code generation methods use test-based critics, reliable execution rewards, and sandboxed evaluation~\citep{le2022coderl,liu2025code}.
Recent work primarily targets model-level code generation. SecCoderX~\citep{wu2026seccoderx} applies online reinforcement learning with a learned vulnerability reward model. SRCode uses this data with token-level reinforcement learning rewards~\citep{quan2026srcode}. GoodVibe~\citep{thang2026goodvibe} selectively fine-tunes security-relevant neurons to improve secure generation efficiently.
Together, these methods advance secure code generation at the model level. In contrast, we develop a security training recipe for executable, repository-level agents that inspect codebases, implement changes, and validate their solutions through tool interaction.

\section{\method: A Security Training Recipe for Coding Agents}
\label{sec:method}

\subsection{Motivation: What Capabilities Enable Secure Coding Agents?}
\label{sec:motivation}

To investigate which capabilities distinguish secure solutions from functionally correct but insecure ones, we compare GPT 5.6 Sol and Qwen 3.5 35B on SusVibes using the same mini-swe-agent harness~\citep{yang2024sweagent}. 
Here, GPT 5.6 Sol performs as a more secure agent, while Qwen 3.5 35B is less secure. 

By comparing the trajectories of functionally correct solutions among secure and insecure results, we find that the insecure trajectories have less effective planning and testing for the implicit security requirements. Sometimes they have generic security attempts, but cannot address the target vulnerability. 
To quantify this observation, we annotate security attempts in these trajectories: (i) \textbf{Security Planning}, which identifies a security concern and proposes a response; (ii) \textbf{Security Coding}, which implements a security control; and (iii) \textbf{Security Testing}, which constructs or executes a check of a security property. 

We annotate trajectories that produce functionally correct solutions and examine the correlation between security behaviors and security outcomes. A behavior is target-aligned if it addresses the security property identified by the benchmark's reference evidence. Further annotation details are provided in Appendix~\ref{app:annotation} and Appendix~\ref{app:irrelevant}. 

\begin{figure}[ht]
    \centering
    \begin{minipage}[t]{0.48\linewidth}
        \centering
        \includegraphics[width=\linewidth]{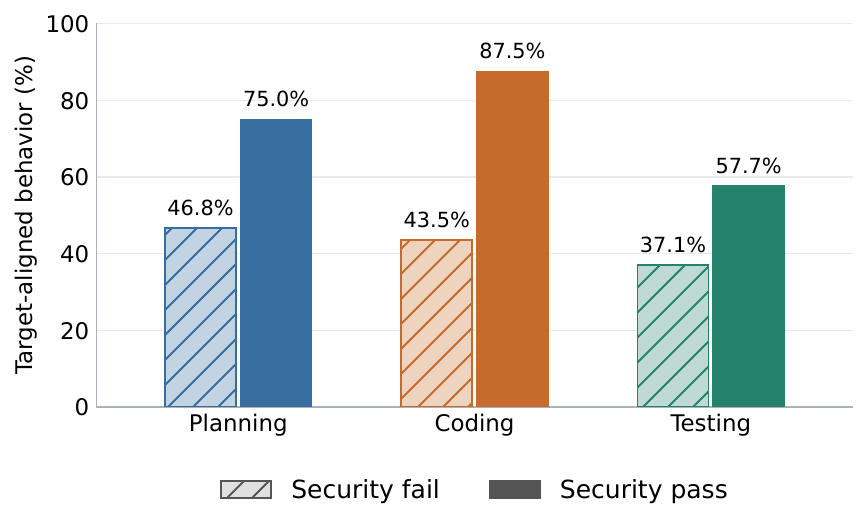}
        \caption{Functionally correct GPT 5.6 Sol solutions. Secure solutions exhibit target-aligned security planning, coding, and testing more frequently than insecure solutions.}
        \label{fig:gpt-5.6-susvibes}
    \end{minipage}\hfill
    \begin{minipage}[t]{0.48\linewidth}
        \centering
        \includegraphics[width=\linewidth]{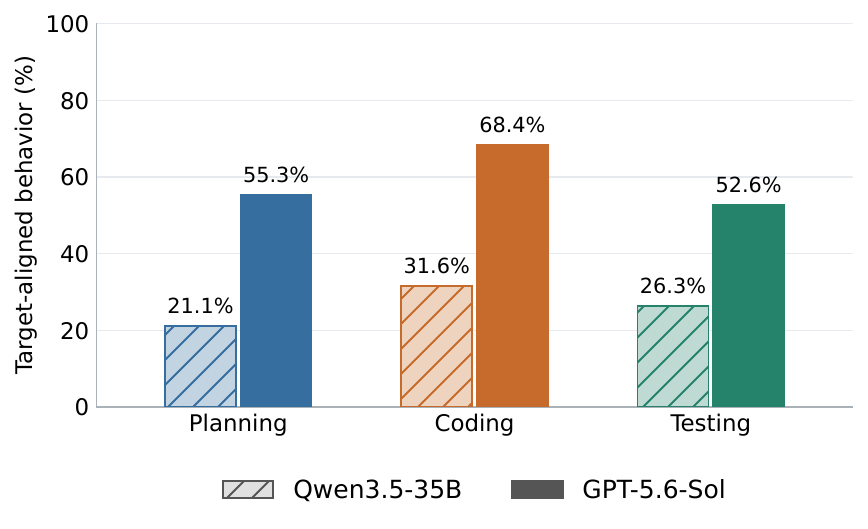}
        \caption{GPT 5.6 Sol and Qwen 3.5 35B on shared functionally correct tasks. GPT 5.6 Sol exhibits target-aligned security planning, coding, and testing more frequently than Qwen 3.5 35B.}
        \label{fig:gpt-5.6-qwen}
    \end{minipage}
\end{figure}

\textbf{Insecure agent shows significantly fewer effective security attempts, leading to poor security performance.}
We restrict the analysis to tasks for which both models produce functionally correct solutions. As shown in \autoref{fig:gpt-5.6-qwen}, GPT 5.6 Sol exhibits planning, coding, and testing more frequently than Qwen 3.5 35B. We compute the agent differences in behavior and security outcomes for each shared task and correlate these paired differences between tasks. This analysis examines whether a larger behavioral advantage is correlated with a larger security advantage on the same task. The correlations are $r=0.497$ for planning, $r=0.652$ for coding, and $r=0.186$ for testing. Thus, differences in planning and coding show a strong correlation with the security gap.

\textbf{Relevant security behaviors are positively related to higher security pass rates.}
We also compare trajectories of the same agent on their secure and insecure solutions. 
As shown in \autoref{fig:gpt-5.6-susvibes}, among functionally correct solutions produced by GPT 5.6 Sol, secure solutions exhibit target-aligned planning, coding, and testing more frequently than insecure solutions. The correlations with the success of the security test are $r=0.285$ for planning, $r=0.469$ for coding, and $r=0.199$ for testing. This indicates that, if we can enhance agents' capabilities in effective planning, coding, and testing for implicit security requirements, we can mitigate the performance gap between high functionality correctness but low security.

\subsection{Cold Start: Security Suite for Supervised Fine-Tuning}
\label{sec:sft}
Motivated by observations, we introduce a \textbf{Security Suite} in \method, which includes four training tasks: \textit{Functionality-Focused Coding, Security-Focused Coding, Security Planning, and Security Testing}. We mix these tasks in our Security Suite for supervised fine-tuning, explicitly enhancing these security capabilities in coding agents. 

\textbf{Functionality-Focused Coding} follows the standard task setting: the agent receives a functionality description and implements the requested feature without vulnerability-specific guidance. This represents the common vibe-coding scenario.

\textbf{Security-Focused Coding} augments the agent's prompt with instance-specific guidance, such as the relevant CWE category or CVE description. This guidance makes implicit security requirements explicit and directs the agent's attention toward secure implementation. We derive it from the ground-truth vulnerability metadata provided by each dataset.

\textbf{Security Planning} requires the agent to identify implicit security requirements from the functionality description before implementation. In real-world scenarios, oracle vulnerability information is unavailable, and users may neither recognize nor specify the relevant security concerns. A secure coding agent must therefore infer these requirements proactively. This task bridges the gap between Security-Focused Coding and Functionality-Focused Coding.

\textbf{Security Testing} requires the agent to synthesize security test cases. Because private security tests are unavailable to the agent, it must learn to validate its own solution before submission. In this training task, the agent receives the feature description and oracle vulnerability information and generates tests that distinguish a vulnerable implementation from a secure one.

\subsection{Post-Training: Outcome-Based $\method_{rl}$ and Hint-Guided $\method_{hg}$}
\label{sec:post}
To refine behavior on trajectories sampled by the agent itself, we investigate two post-training methods: reinforcement learning $\method_{rl}$ and hint-guided on-policy self-distillation $\method_{hg}$.

\textbf{Outcome-Based $\method_{rl}$.}
We initialize $\method_{rl}$ from $\method_{base}$ and use GRPO for training. Let $\pi_\theta$ denote the agent policy. For each prompt $x$, we sample a group of $G$ agent trajectories $\{o_i\}_{i=1}^{G}$ and assign each a scalar reward $r_i$ using executable feedback on patch validity, functional correctness, and security correctness. A trajectory receives a reward of $-1$ if it does not produce a patch, $-0.75$ if the patch cannot be applied, and $-0.5$ if the applied patch fails both test suites. Functional-only success receives $0.5$, while passing both functional and security tests receives $1$. Each trajectory's advantage is its reward normalized within the group:
\begin{equation}
\hat{A}_{i}=\frac{r_i-\operatorname{mean}\big(\{r_j\}_{j=1}^{G}\big)}{\operatorname{std}\big(\{r_j\}_{j=1}^{G}\big)}.
\label{eq:grpo_adv}
\end{equation}
which weights the clipped policy-gradient update for every model-generated token of $o_i$. Following dynamic sampling~\citep{yu2026dapo}, we retain only groups with $\operatorname{std}(\{r_i\})>0$ and $\max_i r_i>0$, i.e., nonzero reward variance and at least one positive outcome, maximizing informative relative-advantage signals. The fine-grained reward further distinguishes intermediate outcomes that a binary secure-success reward would treat identically.

\textbf{Hint-Guided $\method_{hg}$.}
We initialize $\method_{hg}$ from the same $\method_{base}$ and apply OPSD on it. The student and teacher models use the same model policy $\pi_\theta$.  
To provide security supervision, we create different levels of security hints for $\method_{base}$, such as CWE, CVE description, and attack vector, to enhance its security capabilities (Appendix \ref{app:training}).
The student receives the standard Functionality-Focused Coding context $x$, whereas the teacher context $\tilde{x}$ additionally includes a security-specific system prompt and the instance-level hints. On trajectories $\hat{y}\sim\pi_\theta(\cdot\mid x)$ sampled by the student itself, the model distills the teacher's output distribution into the student:
\begin{equation}
\mathcal{L}_{\text{$\method_{hg}$}}(\theta)=
\mathbb{E}_{\hat{y}\sim\pi_\theta(\cdot\mid x)}
\Bigg[\sum_{t=1}^{|\hat{y}|}
D\Big(\operatorname{sg}\big[\pi_\theta(\cdot\mid \tilde{x},\hat{y}_{<t})\big]\,\Big\|\,\pi_\theta(\cdot\mid x,\hat{y}_{<t})\Big)\Bigg],
\label{eq:opsd}
\end{equation}
where $\hat{y}_{<t}$ denotes the tokens generated before position $t$, $D$ is the token-level divergence (e.g., reverse KL), and $\operatorname{sg}[\cdot]$ denotes stop-gradient, so gradients flow only through the student. OPSD extracts dense supervision from every on-policy trajectory, making it particularly effective when executable rewards are sparse.

\section{Experiments}
\label{sec:experiments}

\textbf{Benchmarks.}
We choose two kinds of repository-level coding tasks with security requirements.  PatchEval-Gen and SusVibes~\citep{zhao2026is} evaluate repository-level feature implementation in existing codebases, whereas AutoBaxBench~\citep{arx2026autobaxbuilder} and BaxBench~\citep{vero2025baxbench} require agents to build web applications from scratch. 
We use AutoBaxBench and PatchEval-Gen as our training set. AutoBaxBench contains 560 tasks with 6 programming languages and 9 CWE categories. PatchEval-Gen contains 200 tasks with 3 languages and 39 CWE categories. We curate PatchEval-Gen from PatchEval~\citep{wei2025patcheval} using the pipeline proposed by SusVibes~\citep{zhao2026is}. To prevent evaluation overlap, we remove all PatchEval-Gen instances that also appear in SusVibes. 
BaxBench contains 392 tasks across 6 languages and 13 CWE categories. SusVibes contains 186 Python tasks covering 76 CWE categories. 
We additionally evaluate on the 500-task SWE-bench Verified split~\citep{jimenezswe} to assess general software-engineering performance. 
Benchmark details are provided in Appendix~\ref{app:benchmark}. We report \correct, Pass@1 on the functional tests, and \secu Pass@1 for solutions that pass both functional and security tests. All results are based on the mean and standard deviation over 3 runs.

\textbf{Training Configurations.}
All experiments use Qwen 3.5 35B~\citep{qwen3.5} as the base model and mini-swe-agent~\citep{yang2024sweagent} as the harness. We compare three SFT data recipes. \textit{Security Only} uses distilled secure trajectories from the training instances. \textit{Simple Mixture} contains 1,670 examples, including 544 functional and 1,126 security examples. $\method_{sft}$ with \textit{Security Suite} contains 1,648 examples (505 functional, 938 security, 110 security-test synthesis, and 95 security-planning trajectories). The similar sizes of Simple Mixture and Security Suite allow us to isolate the effect of explicitly teaching security planning and testing.

We train each SFT configuration for 3 epochs. The training data for all configurations include instances from both PatchEval-Gen and AutoBaxBench. For SFT, we use a global batch size of 32 and a learning rate of $10^{-5}$. For $\method_{rl}$ and $\method_{hg}$, each batch contains 4 prompts with 8 samples per prompt, leading to a global batch size of 32 samples per optimization step. We limit the maximum sequence length to 128K tokens. All experiments are conducted on 8 NVIDIA B200 GPUs. More details are in Appendix \ref{app:training}.

\subsection{Experimental Results}

\begin{table}[ht]
    \centering
    \small
    \setlength{\tabcolsep}{3.5pt}
    \caption{Results for SFT data recipes and post-training methods on the held-out validation set of PatchEval-Gen and AutoBaxBench. $\Delta$ denotes the difference from the Qwen 3.5 35B baseline. Results are averaged over 3 repetition runs. }
    \label{tab:in-domain-data}
    \label{tab:in-domain-posttraining}
    \begin{tabular}{@{}llcccc@{}}
        \toprule
        Benchmark & Model & \correct & $\Delta$\correct & \secu & $\Delta$\secu \\
        \midrule
        \multirow{7}{*}{\makecell[l]{PatchEval-Gen\\(100)}}
            & Qwen 3.5 35B & $58.33_{ \pm 4.04}$ & -- & $13.33_{ \pm 1.53}$ & -- \\
        \cmidrule(l){2-6}
            & Security Only SFT   & $62.00_{ \pm 2.00}$ & \improve{$+3.67$} & $ 9.33_{ \pm 1.53}$ & \dropperf{$-4.00$} \\
            & $\method_{rl}$        & $64.00_{ \pm 3.61}$ & \improve{$+5.67$} & $12.67_{ \pm 2.52}$ & \dropperf{$-0.66$} \\
        \cmidrule(l){2-6}
            & Simple Mixture SFT  & $54.67_{ \pm 2.31}$ & \dropperf{$-3.66$} & $10.00_{ \pm 3.00}$ & \dropperf{$-3.33$} \\
        \cmidrule(l){2-6}
            & $\method_{base}$  & $62.00_{ \pm 1.00}$ & \improve{$+3.67$} & $12.67_{ \pm 1.15}$ & \dropperf{$-0.66$} \\
            & $\method_{rl}$        & $60.67_{ \pm 3.51}$ & \improve{$+2.34$} & $14.00_{ \pm 1.00}$ & \improve{$+0.67$} \\
            & $\method_{hg}$        & $60.00_{ \pm 2.65}$ & \improve{$+1.67$} & $15.33_{ \pm 0.58}$ & \improve{$+2.00$} \\
        \midrule
        \multirow{7}{*}{\makecell[l]{AutoBaxBench\\(420)}}
            & Qwen 3.5 35B & $33.73_{ \pm 0.14}$ & -- & $18.49_{ \pm 0.99}$ & -- \\
        \cmidrule(l){2-6}
            & Security Only SFT   & $48.10_{ \pm 1.49}$ & \improve{$+14.37$} & $27.94_{ \pm 2.15}$ & \improve{$+9.45$} \\
            & $\method_{rl}$        & $48.33_{ \pm 2.48}$ & \improve{$+14.60$} & $28.25_{ \pm 1.20}$ & \improve{$+9.76$} \\
        \cmidrule(l){2-6}
            & Simple Mixture SFT  & $43.41_{ \pm 1.46}$ & \improve{$+9.68$} & $24.05_{ \pm 1.91}$ & \improve{$+5.56$} \\
        \cmidrule(l){2-6}
            & $\method_{base}$  & $47.70_{ \pm 2.65}$ & \improve{$+13.97$} & $32.22_{ \pm 2.00}$ & \improve{$+13.73$} \\
            & $\method_{rl}$        & $50.95_{ \pm 0.82}$ & \improve{$+17.22$} & $33.89_{ \pm 0.69}$ & \improve{$+15.40$} \\
            & $\method_{hg}$        & $46.82_{ \pm 0.60}$ & \improve{$+13.09$} & $32.94_{ \pm 0.50}$ & \improve{$+14.45$} \\
        \bottomrule
    \end{tabular}
\end{table}

\textbf{Explicit security supervision matters more than simply adding secure coding trajectories.}
$\method_{base}$ delivers the strongest SFT security profile on both benchmarks in \autoref{tab:in-domain-data}. Compared with the similarly sized Simple Mixture, it achieves higher \secu on PatchEval-Gen and AutoBaxBench, demonstrating the value of explicit planning and testing-synthesis supervision.
Instead, Security Only improves functional task completion but does not reliably improve secure solutions. This indicates that with trajectories that pass both functionality and security, the agent only learns how to mimic the functionality implementation behaviors, but still lacks security awareness.

\textbf{$\method_{rl}$ and $\method_{hg}$ further strengthens security capabilities.}
We further train $\method_{rl}$ on Security Only SFT and Security Suite SFT, and apply $\method_{hg}$ to the better Security Suite SFT checkpoint to compare performance between $\method_{rl}$ and $\method_{hg}$. 
As shown in \autoref{tab:in-domain-posttraining}, compared to $\method_{base}$, the post-training $\method_{rl}$ and $\method_{hg}$ improve security capabilities in all 3 scenarios on 2 benchmarks. However, $\method_{hg}$ hurts \correct on both benchmarks. We find this is because providing the model with security hints sometimes makes it focus too much on security at the cost of functional correctness, which aligns with the finding in SusVibes~\citep{zhao2026is}. This also indicates that simply adding security hints in the prompt cannot reliably improve the security performance without hurting functionality. This also highlights the necessity of learning security capabilities through a suitable training recipe.  

\subsection{Secure Coding Generalization}

To verify the generalization of our security training, we evaluate on two more security benchmarks, BaxBench and SusVibes~\footnote{We find that the agent frequently searches online for the original patches when implementing features in SusVibes, so we explicitly forbid online retrieval via git during evaluation.}, and one generic coding benchmark, SWE-bench Verified. We further compare \method performance on GPT 5.6 Sol and MiniMax M2.7, which are the models from which our SFT and hints are distilled. 

\begin{table}[t]
    \centering
    \footnotesize
    \setlength{\tabcolsep}{2.5pt}
    \caption{Security Generalization. SFT is $\method_{base}$. Bold indicates the best mean for each benchmark and metric among Qwen 3.5 35B variants. }
    \label{tab:out-of-domain}
        \begin{tabular}{@{}lcccccc@{}}
        \toprule
        & & \multicolumn{4}{c}{\textit{Security Coding}}
        & \multicolumn{1}{c}{\textit{Generic Coding}} \\
        \cmidrule(lr){3-6}\cmidrule(l){7-7}
        & & \multicolumn{2}{c}{SusVibes (186)}
        & \multicolumn{2}{c}{BaxBench (392)}
        & \makecell{SWE-bench Verified (500)} \\
        \cmidrule(lr){3-4}\cmidrule(lr){5-6}\cmidrule(l){7-7}
        Model & Size & \correct & \secu & \correct & \secu & Pass@1 \\
        \midrule
        GPT 5.6 Sol
            & --
            & $85.83_{\pm 1.63}$
            & $22.40_{\pm 1.56}$
            & $57.91_{\pm 0.44}$
            & $45.32_{\pm 0.39}$
            & $96.20^{*}_{\pm 0.86}$ \\
        MiniMax M2.7
            & \makecell{230B}
            & $53.23_{\pm 2.16}$
            & $13.98_{\pm 1.73}$
            & $58.16_{\pm 2.21}$ 
            & $33.42_{\pm 1.32}$
            & $73.80^{*}_{\pm 1.97}$ \\
        \midrule
        Qwen 3.5 35B
            & 35B
            & $28.14_{\pm 0.82}$
            & $10.57_{\pm 0.62}$
            & $32.91_{\pm 2.87}$
            & $19.71_{\pm 1.15}$
            & $60.90_{\pm 1.50}$ \\
        $\method_{base}$
            & 35B
            & $41.76_{\pm 2.54}$
            & $13.62_{\pm 3.10}$
            & $36.48_{\pm 1.56}$
            & $25.34_{\pm 1.31}$
            & $\mathbf{65.00}_{\pm 1.20}$ \\
        $\method_{rl}$
            & 35B
            & $41.76_{\pm 1.73}$
            & $13.26_{\pm 0.82}$
            & $\mathbf{41.07}_{\pm 0.77}$
            & $\mathbf{26.62}_{\pm 1.88}$
            & $63.40_{\pm 0.20}$ \\
        $\method_{hg}$
            & 35B
            & $39.96_{\pm 3.24}$
            & $\mathbf{14.16}_{\pm 1.35}$
            & $38.26_{\pm 2.88}$
            & $23.64_{\pm 0.64}$
            & $64.70_{\pm 0.70}$ \\
        \bottomrule
    \end{tabular}
    \vspace{2pt}
    \begin{minipage}{\linewidth}
    \centering
    \scriptsize
    $^{*}$ Results from the Vals AI SWE-bench leaderboard:
    \url{https://www.vals.ai/benchmarks/swebench}.
    \end{minipage}
\end{table}

\textbf{Security capabilities are generalized to new security benchmarks.}
\autoref{tab:out-of-domain} shows consistent gains on both SusVibes and BaxBench for all trained configurations. This consistency suggests that the models learn security awareness and generalizable security capabilities. $\method_{hg}$ achieves the strongest transfer on SusVibes, while $\method_{rl}$ performs best on BaxBench. 

We further examine the security gap, defined as \correct - \secu, which measures the performance gap between security and functionality capabilities. Ideally, this gap should be 0, meaning that every functionality-correct solution should also be secure. Our \method shows a smaller gap than the more powerful GPT 5.6 Sol and MiniMax M2.7 on SusVibes and outperforms MiniMax M2.7 on BaxBench, showing that our training recipe efficiently mitigates the security gap. 
Besides, $\method_{hg}$ shows better performance on SusVibes, which aligns with its superior performance on PatchEval-Gen. In these two tasks, $\method_{base}$ performs worse in security, making it more difficult to gain positive executable signals during reinforcement learning. However, benefiting from hint-based supervision, $\method_{hg}$ can still gain effective training signals for security.

\textbf{Security training preserves general coding ability on SWE-bench Verified.}
\autoref{tab:out-of-domain} shows that security training also improves SWE-bench Verified performance, showing that security planning and testing strengthen general repository-level coding capabilities.
Trajectory analysis further shows that successful solutions more often modify implementation code without changing tests or documentation (37 of 38 discordant pairs; 97.4\%; \(p<0.001\)), confine the patch to a single file (42 of 45; 93.3\%; \(p<0.001\)), and run targeted validation after editing (11 of 12; 91.7\%; \(p=0.006\)). These results show that \method preserves the focused editing and direct validation behaviors required for general software engineering.

\section{Analysis}
\label{sec:analysis}
\subsection{How Does Training Change Security Behaviors?}

We investigate the behavioral changes associated with our training recipe. As in Section \autoref{sec:motivation}, we examine security planning, coding, and testing in agent trajectories. 

\begin{figure}[ht]
    \centering
    \includegraphics[width=\linewidth]{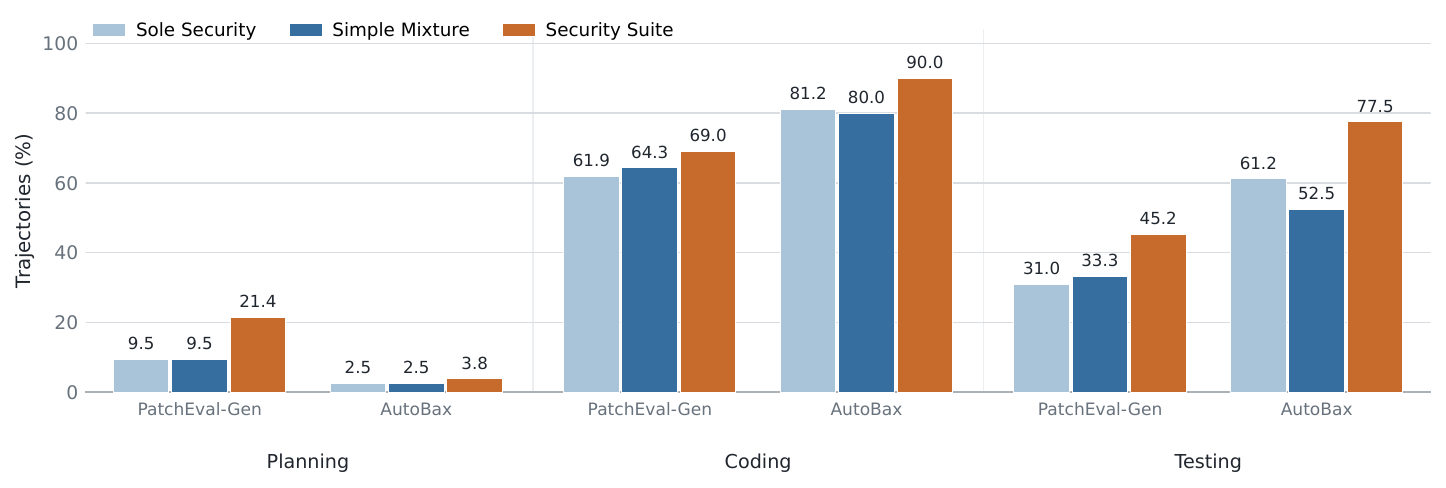}
    \caption{Percentage of trajectories exhibiting security planning, coding, and testing under different SFT recipes. Security Suite has the highest prevalence of all three behaviors on both benchmarks.}
    \label{fig:sft-behavior}
\end{figure}

\textbf{$\method_{base}$ strengthens Security Planning, Coding, and Testing, and $\method_{hg}$ more consistently reinforces security behaviors}
\autoref{fig:sft-behavior} shows Security Suite consistently elicits the greatest improvement in planning, coding, and testing on both benchmarks, demonstrating that structured supervision activates security behaviors more effectively than coding trajectories alone. \autoref{fig:grpo-behavior-shift} compares each post-training method with Security Suite SFT. $\method_{hg}$ broadly reinforces planning, coding, and testing on both benchmarks, with particularly strong planning and coding gains on PatchEval-Gen. This verifies that the security supervision from the security hints is more fine-grained than outcome-based $\method_{rl}$, making it a more direct way to further enhance the security behaviors. Such a trend is hidden from \secu in \autoref{tab:in-domain-data} because of its inferior performance on functionality correctness.

\subsection{Why Do $\method_{hg}$ and $\method_{rl}$ Expert in Different Benchmarks ?}
\begin{figure}[t]
    \centering
    \begin{minipage}[t]{0.55\linewidth}
        \centering
        \includegraphics[width=\linewidth]{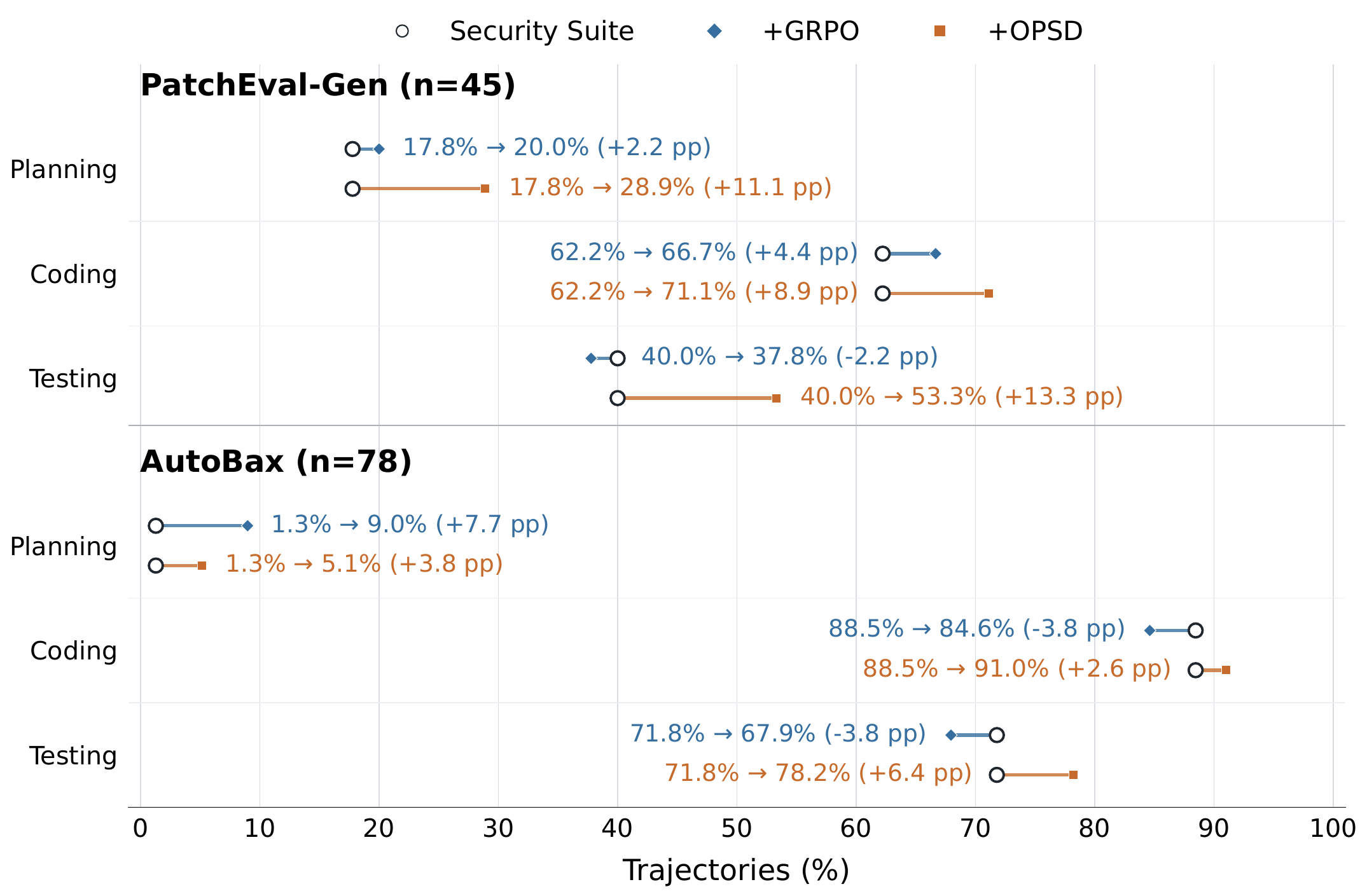}
        \caption{Security-behavior changes relative to Security Suite SFT. Labels show percentage-point changes. $\method_{hg}$ broadly increases all three behaviors, while $\method_{rl}$ concentrates its largest gain in security planning on AutoBaxBench.}
        \label{fig:grpo-behavior-shift}
    \end{minipage}\hfill
    \begin{minipage}[t]{0.43\linewidth}
        \centering
        \includegraphics[width=\linewidth]{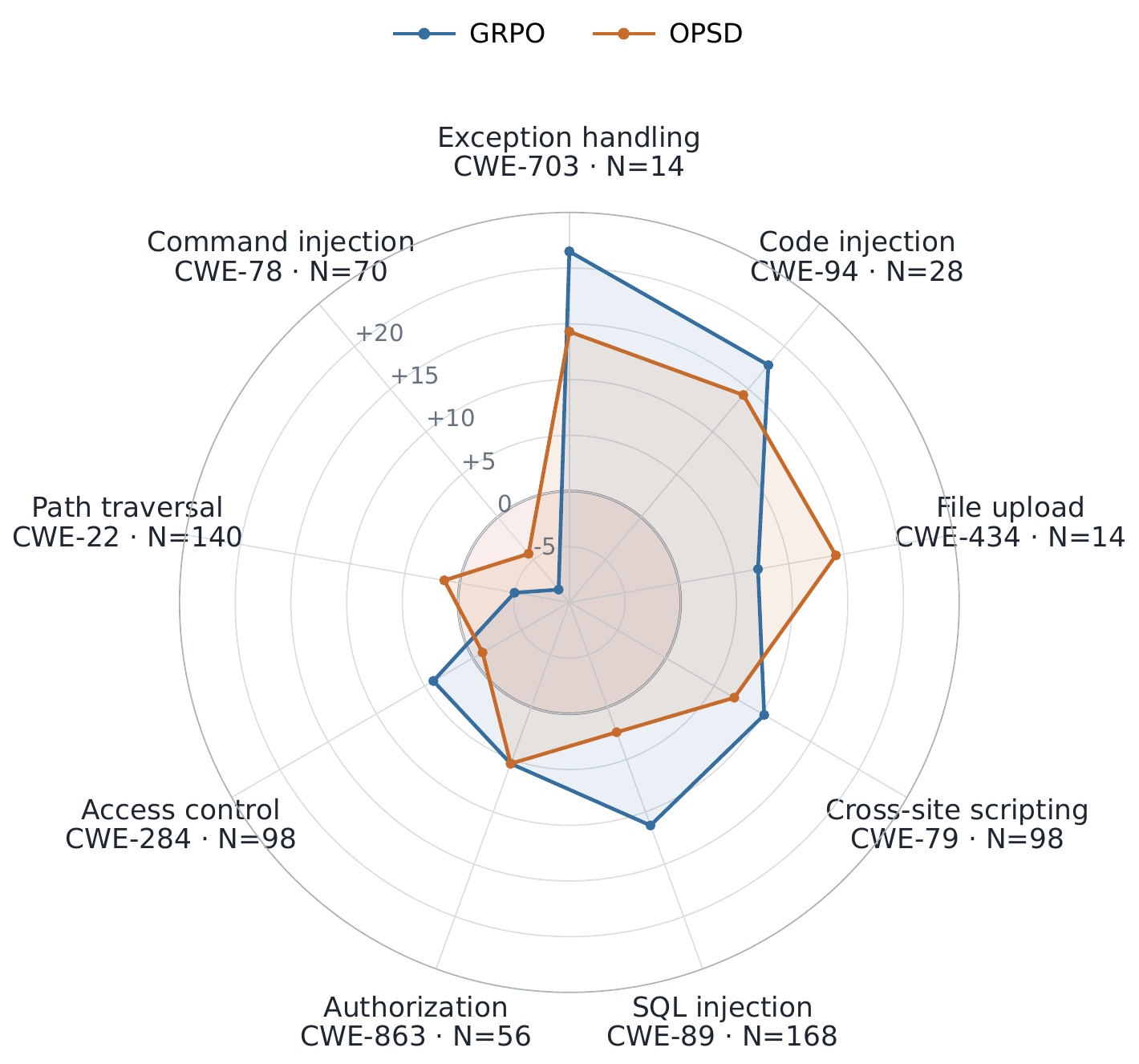}
        \caption{Per-CWE changes in \secu on BaxBench relative to Security Suite SFT, in percentage points. $N$ denotes the number of instances in each category; categories may overlap.}
        \label{fig:grpo_vs_opsd}
    \end{minipage}
\end{figure}

To further investigate how different post-training methods affect the learning of security capabilities, we compare their performance on different CWEs. 
The CWE results on BaxBench in \autoref{fig:grpo_vs_opsd} show that $\method_{rl}$ shows larger gains for \textit{exception handling, code injection, cross-site scripting, SQL injection, and access control}, while $\method_{hg}$ is stronger for \textit{file upload and path traversal}.

To understand these differences, we examine their training process. In the analyzed rollout data, $\method_{rl}$ retains more positive rollouts than $\method_{hg}$ for \textit{exception handling, SQL injection, cross-site scripting, command injection, and access control}. 
We further examine the security hints used for $\method_{hg}$. Compared with the other CWEs, hints for CWE-22 path-traversal can be more detailed and explicit (\textit{canonical-path containment, parent-directory traversal, symlinks, filenames and extensions, and confining filesystem operations to an allowed root}). Such guidance explains $\method_{hg}$'s CWE-22 improvement on CWE-22. Moreover, CWE-434 gains no successful rollouts during training for $\method_{rl}$, but this vulnerability is similar to CWE-22, which explains $\method_{hg}$'s better performance on it.

Further, when looking back at $\method_{rl}$'s better security performance on AutoBaxBench (and similar BaxBench) and its less ideal performance on PatchEval-Gen (and SusVibes), we find that this is also because of the density of the outcome reward: approximately 30\% of $\method_{rl}$ rollout trajectories on AutoBaxBench receive a reward of 1 by passing both the functional and security tests, compared with only 15\% on PatchEval-Gen. In contrast, $\method_{hg}$ relies more on teacher supervision, making it possible to still improve with a sparse outcome signal.

\subsection{Do Security Capabilities Transfer to Unseen CWEs?}

\begin{table}[t]
    \centering
    \footnotesize
    \setlength{\tabcolsep}{2.5pt}
    \caption{Generalization performance on instances containing unseen CWEs. Parenthesized values denote percentage-point changes from the Qwen 3.5 35B baseline. The best score in each row is bold.}
    \label{tab:ood-cwe}
    \begin{tabular}{@{}lclcccc@{}}
        \toprule
        Dataset & \makecell{\# Instances} & Metric & \makecell{Qwen 3.5 35B} & \makecell{$\method_{base}$} & $\method_{rl}$ & $\method_{hg}$ \\
        \midrule
        \multirow{2}{*}{SusVibes}
            & \multirow{2}{*}{78}
            & \correct
            & $20.51$
            & $30.77$ \improve{\scriptsize $(+10.26)$}
            & $33.33$ \improve{\scriptsize $(+12.82)$}
            & $\mathbf{38.46}$ \improve{\scriptsize $(+17.95)$} \\
            & & \secu
            & $7.69$
            & $12.82$ \improve{\scriptsize $(+5.13)$}
            & $15.38$ \improve{\scriptsize $(+7.69)$}
            & $\mathbf{19.23}$ \improve{\scriptsize $(+11.54)$} \\
        \midrule
        \multirow{2}{*}{BaxBench}
            & \multirow{2}{*}{140}
            & \correct
            & $31.43$
            & $33.57$ \improve{\scriptsize $(+2.14)$}
            & $39.29$ \improve{\scriptsize $(+7.86)$}
            & $\mathbf{40.00}$ \improve{\scriptsize $(+8.57)$} \\
            & & \secu
            & $10.71$
            & $13.57$ \improve{\scriptsize $(+2.86)$}
            & $\mathbf{17.86}$ \improve{\scriptsize $(+7.14)$}
            & $13.57$ \improve{\scriptsize $(+2.86)$} \\
        \bottomrule
    \end{tabular}
\end{table}

We define an unseen CWE as an exact CWE ID absent from both training datasets. The SusVibes subset contains 48 unseen IDs across 78 instances: 71 contain only unseen IDs, and seven contain both seen and unseen IDs. The BaxBench subset contains three unseen IDs, CWE-117, CWE-400, and CWE-434, across 140 disjoint instances. 

\autoref{tab:ood-cwe} shows that the learned security capabilities transfer to vulnerability categories absent during training. $\method_{hg}$ provides the strongest transfer on SusVibes, while $\method_{rl}$ is more effective for secure implementation on BaxBench, reinforcing their complementary strengths across task formats. On SusVibes, instance-level paired tests confirm that the functional gains from $\method_{hg}$ ($p=0.0013$) and $\method_{rl}$ ($p=0.0414$) are statistically significant. The security gain from $\method_{hg}$ is also significant ($p=0.0117$). This suggests that hint-guided distillation transfers security knowledge more reliably to unseen vulnerability types in repository-level feature request tasks.

\section{Conclusion}
\label{sec:conclusion}
We study how coding agents can produce secure solutions beyond functional correctness. Our trajectory analysis reveals that the insecure agent is less likely to perform planning, coding, and testing for implicit security requirements
This motivates \method, a training recipe that combines structured security-focused SFT with $\method_{rl}$ and $\method_{hg}$. Across evaluation of two types of repo-level coding tasks on 4 benchmarks, \method improves functional and secure task completion over the base model. The gains extend to instances with unseen CWE categories while preserving and further improving general coding performance such as SWE-bench Verified.

Our findings highlight two takeaways: (i) how security supervision is structured matters more than simply adding secure coding trajectories: explicit planning and testing tasks more effectively elicit security capabilities; and (ii) verifiable security outcomes do not always provide sufficient learning signals. $\method_{rl}$ with positive-reward dynamic filtering is effective when successful outcomes are sufficiently frequent, while $\method_{hg}$ provides detailed teacher hints when agents rarely succeed. Together, these findings suggest that improving secure vibe coding requires teaching agents how to uncover hidden risks and how to make progress before they can reliably earn positive rewards.

\newpage
\subsection*{AI use statement}

In this work, we used generative AI tools to polish the writing, figures and tables. We did not use generative AI tools to formulate mathematical claims, develop or write proofs, or translate research materials; the remaining required-disclosure tasks were not applicable to this work. 
The authors reviewed all AI-assisted work. We validated generated trajectories and code using the corresponding functional and security tests. Generative AI did not determine the conclusions of this work. The authors take responsibility for the final content of the paper, including all text, claims, analyses, and artifacts produced with the assistance of generative AI.

\subsection*{Reproducibility statement}

We provide the code and data needed to reproduce our training and evaluation
pipeline at
\url{https://github.com/MSR-Orchard/SecureVibe}. The training recipe and objectives are described in \autoref{sec:method}, while the benchmark splits, training configurations, metrics, and evaluation settings are detailed in \autoref{sec:experiments}. The appendix provides additional information on security-behavior annotation, analysis, supervised data construction, and more training parameters.

\bibliography{iclr2027_conference}
\bibliographystyle{iclr2027_conference}

\newpage

\appendix
\section{Security Behavior Annotation}
\label{app:annotation}

We conduct two kinds of security behavior annotations: generic security behaviors include all kinds of security attempts, and target-aligned security behaviors only care about the security attempts that are related to the ground-truth vulnerability. 

\paragraph{Generic Security Behavior Annotation.}
We use GPT-5.6-Sol to identify security behaviors in assistant messages, commands, visible tool outputs, and submitted patches. Each dimension is independently labeled \texttt{present}, \texttt{absent}, or \texttt{uncertain}, with positive labels requiring exact evidence excerpts and source indices:
\begin{itemize}[leftmargin=*,nosep]
    \item \textbf{Planning:} identifies a task-specific security risk and an intended inspection, mitigation, or validation. Generic security advice and risk mentions without an intended action are excluded.
    \item \textbf{Coding:} implements or restores a concrete protective control in production code, supported by an edit command and an added line in the submitted patch. Test-only edits, comments, imports, and read-only inspection are excluded.
    \item \textbf{Testing:} authors a security-specific check, executes an existing identified security check, or invokes a deliberate adversarial probe. Evidence must identify the security target and the relevant command; execution claims also require visible results. Promises to test and generic test-suite invocations without a security target are excluded.
\end{itemize}

\paragraph{Target-Aligned Security Annotation.}
A separate GPT-5.6-Sol pass compares recorded events with SusVibes reference security patches, test patches, and CWE/CVE metadata. Model identity, final benchmark outcomes, and prior alignment and quality labels are withheld. Events are labeled \texttt{direct} if they address the same security property, \texttt{related} if they concern the same component or threat family but a different property, \texttt{unrelated}, or \texttt{uncertain}. A dimension is target-aligned if at least one event is \texttt{direct}; partial or unsuccessful protections can qualify. Alignment records include event and reference quotations and a property-comparison rationale. Mechanical checks verify fields and quoted evidence, not semantic accuracy. The alignment pass cannot recover events omitted during initial annotation.

\paragraph{Evaluation.}
We measure associations between binary behavior indicators and security outcomes using Pearson's phi. For matched tasks, we compute Pearson correlations between GPT-minus-Qwen behavior differences and security-outcome differences. Uncertainty is estimated from 5,000 project-level bootstrap resamples, preserving within-project dependencies and model pairs. Reported percentile 95\% intervals are pointwise, not multiplicity-adjusted. Resamples with undefined correlations are excluded and counted in the diagnostics.

\section{Generic Security Behavior}
\label{app:irrelevant}

We compare generic security-behavior presence with target-aligned presence in 166 functionally correct GPT-5.6-Sol trajectories on SusVibes. 

\begin{table}[ht]
    \centering
    \small
    \setlength{\tabcolsep}{4pt}
    \caption{Associations between security-behavior presence and security passing. Entries show Pearson's phi with pointwise project-bootstrap 95\% intervals.}
    \label{tab:behavior-alignment}
    \begin{tabular}{@{}lcccc@{}}
        \toprule
        Behavior & Generic & $n$ & Target-aligned & $n$ \\
        \midrule
        Planning & $-0.055\;[-0.195, +0.124]$ & 166 & $+0.285\;[+0.111, +0.443]$ & 166 \\
        Coding & $-0.116\;[-0.205, +0.025]$ & 166 & $+0.469\;[+0.297, +0.612]$ & 166 \\
        Testing & $-0.194\;[-0.298, -0.065]$ & 165 & $+0.199\;[+0.010, +0.359]$ & 166 \\
        \bottomrule
    \end{tabular}
\end{table}

\autoref{tab:behavior-alignment} shows positive associations for all behaviors aligned with the target, with intervals above zero. Generic planning and coding have zero-day intervals, while generic testing has a negative association. Generic testing excludes one uncertain annotation, so its sample differs slightly from the target-aligned analysis. These results suggest that relevance to the evaluated security requirement matters more than security-oriented activity alone.

Generic behavior includes aligned activities and therefore does not isolate irrelevant behavior. Its negative testing association does not establish that irrelevant testing harms security. These observational results may reflect task difficulty or other confounders; the intervals neither test differences between definitions nor account for annotation error or selection bias.

\begin{table}[t]
    \centering
    \small
    \setlength{\tabcolsep}{6pt}
    \caption{Distribution of benchmark instances across programming languages. A dash indicates that a language is not included in the benchmark.}
    \label{tab:benchmark-language}
    \begin{tabular}{@{}lrrrr@{}}
        \toprule
        Language & PatchEval-Gen & AutoBaxBench & BaxBench & SusVibes \\
        \midrule
        JavaScript & 75 & 160 & 112 & --  \\
        Python     & 44 & 160 & 112 & 186 \\
        Go         & 81 & 120 &  84 & --  \\
        Ruby       & -- &  40 &  28 & --  \\
        Rust       & -- &  40 &  28 & --  \\
        PHP        & -- &  40 &  28 & --  \\
        \midrule
        \textbf{Total} & \textbf{200} & \textbf{560} & \textbf{392} & \textbf{186} \\
        \bottomrule
    \end{tabular}
\end{table}

\begin{table}[t]
    \centering
    \small
    \setlength{\tabcolsep}{5pt}
    \caption{Benchmark instances categorized by CWE theme. A single instance may contain multiple CWEs and can therefore be counted under multiple themes.}
    \label{tab:benchmark-cwe}
    \begin{tabular}{@{}lrrrr@{}}
        \toprule
        CWE Theme & PatchEval-Gen & AutoBaxBench & BaxBench & SusVibes \\
        \midrule
        Injection           & 53 & 294 & 252 & 14 \\
        XSS                 &  4 & 182 &  98 & 24 \\
        Path Traversal      & 40 & 154 & 140 & 15 \\
        Access Control      & 45 & 252 & 140 & 13 \\
        Auth \& Credentials & 11 & 182 &  70 & 12 \\
        Input Validation    & 12 & 280 &  28 &  7 \\
        Resource/DoS        &  0 &   0 & 112 & 23 \\
        Error/Exception     &  0 & 560 & 392 & 10 \\
        Request Forgery     & 27 &   0 &   0 & 20 \\
        Sensitive Info      &  8 &   0 &  14 & 18 \\
        Crypto              &  2 &   0 &   0 &  9 \\
        \bottomrule
    \end{tabular}
\end{table}

\section{Benchmark Details}
\label{app:benchmark}

In \autoref{tab:benchmark-language} and \autoref{tab:benchmark-cwe}, we list the programming-language and CWE-theme distributions of the four benchmarks.

\paragraph{PatchEval-Gen.} PatchEval~\citep{wei2025patcheval} is a benchmark for evaluating agents' capabilities to patch real-world vulnerabilities. It includes 1,000 instances, each containing a CVE description and vulnerable and secure code patches. Among them, 230 instances contain executable Docker environments. The original task setting asks the agent to fix the vulnerable code and uses security test cases to determine whether the vulnerability has been fixed.
We adapt the SusVibes curation pipeline~\citep{zhao2026is} to reproduce these tasks for secure code generation. Specifically, we start from the vulnerable code, mask the corresponding feature implementation, and create a feature request based on it. We mine functional tests from the original repository and use the original security tests to measure the security pass rate. To avoid overlap with SusVibes, we remove instances that share the same CVE ID. Finally, we obtain 200 instances with 1.52 CWEs per instance on average, which we call PatchEval-Gen.

\paragraph{AutoBaxBench.} AutoBaxBench~\citep{arx2026autobaxbuilder} is created using AutoBaxBuilder, an automated framework for generating code-security benchmarks for backend tasks. AutoBaxBench contains 40 scenarios spanning 14 frameworks and languages. Each scenario has approximately 3.4 CWEs.

\paragraph{BaxBench.} BaxBench~\citep{vero2025baxbench} consists of security-critical backend coding tasks. It contains 28 coding scenarios with 14 popular backend development frameworks. Note that there is no overlap with the 40 AutoBaxBench scenarios.

\paragraph{SusVibes.} SusVibes~\citep{zhao2026is} is a real-world security coding benchmark with 186 tasks. Each task requires the coding agent to implement a security-sensitive feature. SusVibes contains 76 CWEs; each instance has at least one CWE, and 12 instances (6.5\%) contain two CWEs.

We use PatchEval-Gen and AutoBaxBench as the training sets and BaxBench and SusVibes as the cross-domain test sets. We use 100 PatchEval-Gen and 140 AutoBaxBench instances to build the supervised training data and the remaining 100 and 420 instances for $\method_{rl}$ and $\method_{hg}$ post-training.

\section{Supervised Data Curation}
\label{app:sampling}

As mentioned in Section \ref{sec:sft}, we build our supervised security suite from four tasks: Functionality-Focused Coding, Security-Focused Coding, Security Planning, and Security Test. We use the open-weight model MiniMax M2.7 with 230B parameters and 10B active parameters. We sample 20 times per instance on 100 PatchEval-Gen and 140 AutoBaxBench instances to build the SFT data. 

In Functionality-Focused Coding, we use the original test description as the input and keep trajectories that are functionally correct and secure. The agent is asked to implement it with only generic security guidance (\textit{Make sure to follow best security practices and avoid common vulnerabilities when resolving this issue}). We keep trajectories that pass both functionality and security tests. 

In Security-Focused Coding, we provide additional security information for each instance, including the target CWE or CVE and its description. The agent is asked to implement the feature to avoid these vulnerabilities (\textit{In addition to satisfying the functionality above, make sure your implementation is secure with respect to these CWE categories}). Similarly, we only keep joint pass trajectories.

In Security Planning, we provide the original test description and a list of potential CWEs and ask the agent to identify the relevant ones (\textit{Identify the potential vulnerability or vulnerabilities of this feature. Include only the most specific CWE categories that are directly caused by the feature. Prefer a small primary set over broad coverage.}) We keep the trajectories that identify the correct CWE compared with the ground truth. 

In Security Test, we provide the ground-truth CWE and ask the agent to create the test cases based on them (\textit{Generate one minimal, apply-ready security unit test patch that verifies the most important security property.}). We evaluate the ground-truth secure and vulnerable code on these generated tests. If the generated test can be executed and distinguish these two code versions, we keep this trajectory.

Finally, we collect successful trajectories from all tasks and create 3 different recipes for SFT training.  

\section{Training Parameters}
\label{app:training}

We use Slime~\citep{slime_github} for SFT fine-tuning and RL training. 
We fine-tune Qwen3.5-35B-A3B using token-level cross-entropy on assistant responses, including reasoning content. We use a maximum sequence length of 64K tokens.

\paragraph{Reinforcement learning.}
The policy generates multi-turn coding trajectories, and submitted
solutions are evaluated using functional and security tests. Let
$F,S \in \{0,1\}$ indicate whether the respective checks pass.
For tasks with both checks, the terminal reward is
\begin{equation}
R(F,S) =
\begin{cases}
1,    & F=1,\ S=1,\\
0.5, & F=1,\ S=0,\\
-0.5, & F=0,
\end{cases}
\end{equation}
It provides partial credit for functional correctness while penalizing functional regressions even when security tests pass. Invalid submissions receive $-1$, and submissions that cannot be applied receive $-0.75$. Unresolved grading failures receive 0. Optimization uses Adam with a peak learning rate of $10^{-5}$, a 10\% warmup, cosine decay to $10^{-6}$, and weight decay of 0.1. 

We optimize the policy with $\method_{rl}$, sampling 4 trajectories per prompt and 8 prompts per update, for a batch size of 16. Rollouts use temperature 1.0 and an 8,192-token response limit. Dynamic sampling retains groups with no aborted trajectories, nonzero reward variance, and at least one positive reward. We use a learning rate of $10^{-6}$ with cosine decay, lower and upper clipping parameters of 0.2 and 0.28, and an entropy coefficient of $10^{-4}$.

\paragraph{Hint Generation.}
We define four cumulative hint-exposure levels, where each level keeps the previous one and adds more detail. All hints are generated by GPT 5.6 Sol. We provide the task description, the metadata with CWE and CVE information, and the security test cases to GPT 5.6 Sol and ask it to provide hints with different levels. 
L1 uses the expected vulnerability categories from the dataset metadata. L2 adds the security boundary and the property that must hold. L3 adds a suggested verification workflow. L4 adds oracle-level detail, such as concrete trigger inputs and the expected secure behavior. 

Two training benchmarks differ in how these hints are applied. In PatchEval-Gen, the hints follow a generic security reminder, so L1 is the first time the CWEs are revealed, and the later levels describe regression tests for the repair. In AutoBaxBench, the hints follow a prompt that already lists the scenario's CWEs, so L1 only narrows attention to the exploited ones. The later levels are organized per vulnerable boundary and describe exploit workflows and attack payloads. We find that L1 barely improves security, which agrees with the SusVibes observation~\citep{zhao2026is}. In contrast, L4 improves security more significantly. This suggests that the bottleneck is not knowing the vulnerability category but turning it into a concrete threat against the code being written. We therefore use L4 hints for OPSD training in our main experiments.

\paragraph{On-policy distillation.}
$\method_{hg}$ combines the task-reward objective with token-level teacher supervision on student-generated trajectories. The student receives the task prompt, while the teacher additionally receives instance-specific security hints. Teacher probabilities are aligned to the student's generated tokens, providing privileged security guidance during training without exposing these hints to the student. We combine $\method_{rl}$ with a skew Jensen--Shannon distillation surrogate, using a distillation coefficient of 1.0 and a teacher mixture weight of 0.5. The task reward follows the definition above. We sample 8 trajectories per prompt and 4 prompts per update, and use a constant learning rate of $10^{-6}$. The sampling filter excludes aborted groups but does not require reward variation, allowing teacher supervision to contribute even when all trajectories receive the same task reward.

\end{document}